\documentclass[aps,prb,reprint,nofootinbib,floatfix]{revtex4-2}
\usepackage{amsmath,amssymb,bm}
\usepackage{graphicx}
\usepackage[hidelinks]{hyperref}
\usepackage{color}

\newcommand{\dd}{\mathrm{d}}
\newcommand{\sgn}{\operatorname{sgn}}
\newcommand{\eps}{\varepsilon}
\newcommand{\calC}{\mathcal{C}}
\newcommand{\Aph}{A^{\rm ph}}
\newcommand{\Aem}{A^{\rm em}}

\newcommand{\zhat}{\hat{\mathbf z}}

\begin{document}

\title{
Thermal effect on the anomaly-induced electromechanical response in gapped 
graphene}

\author{A. Sedrakyan$^{1}$,  and K. Ziegler$^{2,3}$}

\affiliation{$^1$Yerevan Physics Institute, Br. Alikhanian 2, Yerevan 36, Armenia,\\
    $^{2}$Physics Department, New York City College of Technology,\\
   The City University of New York, Brooklyn, NY 11201, USA \\
$^3$   Institut f\"ur Physik, Universit\"at Augsburg, D-86135 Augsburg, Germany
}

\begin{abstract}
Mechanical deformation of gapped graphene can act on Dirac quasiparticles as an emergent gauge field.  When this deformation field couples to the same current as the electromagnetic field, the parity anomaly produces a mixed electromechanical Chern-Simons response: a phonon electric field drives a transverse electrical current, and a phonon magnetic field binds charge.  Previous zero-temperature results predict a sharp change in the response when the chemical potential crosses the band edge.  We show that finite temperature replaces this sharp feature by a universal smooth crossover controlled only by the ratios of temperature, gap, and chemical potential.  The response remains almost quantized in the insulating regime, is rounded over a gate window of order temperature near the band edge, and approaches the doped Berry-curvature result with a controlled Sommerfeld correction.  We apply the result to two experimentally useful drives: a traveling flexural wave, which produces a transverse second-harmonic current, and a dynamic phonon mixed with a static ripple, which produces a fundamental-frequency signal.  The same gate-temperature line shape controls both signals.  This gives a direct way to separate the anomaly-induced current from ordinary electromechanical backgrounds and to extract the effective gap and electronic temperature in graphene devices.
\end{abstract}

\maketitle

\section{Introduction}

The low-energy electrons in graphene are described by Dirac fermions near two valleys.  Smooth deformations of the sheet modify this Dirac theory.  They change the local velocity, generate scalar and mass-like terms, and also produce gauge-like fields.  This description is a standard way to organize strain effects in graphene and related two-dimensional Dirac materials~\cite{Vozmediano2010,Amorim2016}.  It is also closely related to the geometric description of Dirac fermions constrained to a curved surface.  In the formulation developed in Refs.~\cite{SinnerSedrakyanZiegler2011,SedrakyanSinnerZiegler2021}, deformations of a graphene sheet generate an emergent phonon gauge field.  For a purely flexural displacement this field starts at quadratic order in the displacement.  As a result, a monochromatic mechanical drive can generate electrical response at a harmonic of the drive frequency.

A recent analysis of deformed gapped graphene identified a particularly simple consequence of this structure~\cite{Sedrakyan2026FiniteFrequency}.  If the deformation-induced gauge field $\Aph_\mu$ couples to the same Dirac current as the electromagnetic field $\Aem_\mu$, the parity-odd part of the current-current response contains a mixed electromagnetic-phonon Chern-Simons term.  
This reflects also the fact that the phonons inherit the chiral properties
of the electronic Dirac Hamiltonian in the form of two touching phonon
bands and an intermediate flat band~\cite{ziegler26}.
In physical terms, a phonon electric field drives a transverse charge current, while a phonon magnetic field induces charge density.  For a single massive two-component Dirac cone, the zero-temperature local coefficient has the Chern-Simons value when the chemical potential lies in the gap and is reduced to the Berry-curvature factor $m/|\mu|$ when the chemical potential lies in a band.  In sublattice-gapped graphene with the valley-odd deformation gauge coupling, the two valleys and the two spin projections add in the charge channel.  The effect is then an electrical response rather than a purely valley response~\cite{Sedrakyan2026FiniteFrequency}.

The next experimentally relevant question is how this response changes at finite temperature.  At zero temperature the local coefficient has a nonanalytic gate dependence at $|\mu|=|m|$.  A real gate sweep in graphene on hBN, suspended graphene, or another gapped Dirac device will not show an infinitely sharp edge, which will show thermal rounding. The latter, as well
as a thermally robust phonon anomaly, was found for phonons~\cite{ziegler26}.
This rounding is not an uncontrolled detail.  In the local, low-frequency regime the coefficient is fixed by the Berry curvature of the massive Dirac bands and by Fermi occupation factors.  The finite-temperature line shape is therefore universal once the gap, chemical potential, and temperature are specified.

This paper derives that universal line shape and applies it to the deformation protocols proposed for anomaly-induced electromechanical response.  The central result is a finite-temperature coefficient $\calC_T(m,\mu)$ that replaces the zero-temperature factors $\sgn(m)$ and $m/|\mu|$.  We give its exact integral representation, its limiting forms in the gap, at the band edge, and in the doped regime, and numerical gate-temperature maps.  We then show how the same coefficient controls a transverse second-harmonic current from a traveling flexural wave and a fundamental-frequency current obtained by mixing a static ripple with a dynamic phonon.

We set $\hbar=k_B=1$.  The microscopic strength of the deformation-induced gauge field is absorbed into a phenomenological coupling $g_{\rm ph}$.  This convention keeps the universal Dirac response separate from the microscopic calibration of strain or flexural deformation in a particular device.

\section{Local mixed Chern-Simons response}

Consider one massive Dirac cone with Hamiltonian
\begin{equation}
H({\bf k})=v(k_x\sigma_x+k_y\sigma_y)+m\sigma_z-\mu .
\label{eq:H}
\end{equation}
The electromagnetic and phonon gauge fields enter through the same current vertex,
\begin{equation}
{\bf k}\rightarrow {\bf k}-e{\bf A}^{\rm em}-g_{\rm ph}{\bf A}^{\rm ph} .
\end{equation}
The parity-odd part of the one-loop effective action then contains the mixed term \cite{Niemi-1983,Redlich1984PRL,Semenoff1984,Jackiw1984}
\begin{equation}
S_{\rm mix}^{\rm odd}=\frac{e g_{\rm ph}}{4\pi}\,\calC_T(m,\mu)
\int \dd t\,\dd^2x\,\epsilon^{\mu\nu\rho} A^{\rm em}_{\mu}\partial_\nu A^{\rm ph}_{\rho} .
\label{eq:Smix}
\end{equation}
This equation fixes the sign convention used below.  Reversing the orientation convention of a Dirac cone, or reversing the sign convention for the deformation gauge field, reverses the overall sign.  The dependence on gate voltage and temperature is unchanged.

Varying Eq.~\eqref{eq:Smix} with respect to $A^{\rm em}_\mu$ gives the local anomalous charge and current,
\begin{align}
\rho_{\rm anom}&=\Gamma_T B^{\rm ph},\label{eq:rho}\\
{\bf j}_{\rm anom}&=\Gamma_T\,\zhat\times {\bf E}^{\rm ph},\label{eq:j}
\end{align}
where
\begin{equation}
\Gamma_T=\frac{e g_{\rm ph}}{4\pi}\calC_T(m,\mu),\quad
B^{\rm ph}=\partial_x A^{\rm ph}_y-\partial_y A^{\rm ph}_x,
\end{equation}
and
\begin{equation}
{\bf E}^{\rm ph}=-\partial_t{\bf A}^{\rm ph}-\nabla A^{\rm ph}_0 .
\end{equation}
Eqs.~\eqref{eq:rho} and \eqref{eq:j} are local linear response laws.  They apply when the mechanical frequency and wave vector are small compared with the electronic scales that control nonlocal corrections.  They also assume that the drive is below the particle-hole continua that produce dissipative interband or intraband absorption.  The complete finite-frequency response at finite density contains more tensor structures because the Fermi sea selects a rest frame.  Here we focus on the low-frequency lock-in limit in which the Hall coefficient is local.

\section{Finite-temperature coefficient}

The coefficient $\calC_T$ follows from the Berry-curvature form of the local anomalous Hall response.  The band energies are $E_\pm({\bf k})=\pm \eps_k$, with
\begin{equation}
\label{Ek}
\eps_k=\sqrt{v^2k^2+m^2}.
\end{equation}
With the orientation convention of Eq.~\eqref{eq:Smix}, the two Berry curvatures are
\begin{equation}
\label{omega}
\Omega_\pm({\bf k})=\mp \frac{m v^2}{2\eps_k^3}.
\end{equation}
The local mixed Hall coefficient is therefore (see Appendix A and reference there)
\begin{align}
\frac{\sigma^{\rm mix}_{xy}}{e g_{\rm ph}/4\pi}
=4\pi\int\frac{\dd^2k}{(2\pi)^2}&\frac{m v^2}{2\eps_k^3}\\
&\times\left[n_F(-\eps_k-\mu)-n_F(\eps_k-\mu)\right], \nonumber
\end{align}
where $n_F(x)=1/(e^{x/T}+1)$. 
The difference of the two Fermi functions originates from the opposite Berry curvatures of the lower and upper bands \cite{Xiao-2010}.
 Changing variables from $k$ to $\eps$ by using 
$\varepsilon d\varepsilon=v k d (v k)$ from (\ref{Ek}) gives
\begin{equation}
\calC_T(m,\mu)=m\int_{|m|}^{\infty}\frac{\dd \eps}{\eps^2}
\left[n_F(-\eps-\mu)-n_F(\eps-\mu)\right].
\label{eq:C_FD}
\end{equation}
For $T>0$ this may also be written as
\begin{equation}
\calC_T(m,\mu)=m\int_{|m|}^{\infty}\frac{\dd \eps}{\eps^2}\,
\frac{\sinh(\eps/T)}{\cosh(\mu/T)+\cosh(\eps/T)} .
\label{eq:C_closed}
\end{equation}
Eq.~\eqref{eq:C_closed} is independent of the velocity $v$.  The velocity only enters the size of nonlocal corrections and the relation between momentum and electronic energy.

Several limits follow directly.  At zero temperature,
\begin{equation}
\calC_0(m,\mu)=
\begin{cases}
\sgn(m), & |\mu|<|m|,\\[3pt]
{m}/{|\mu|}, & |\mu|>|m|.
\end{cases}
\label{eq:T0}
\end{equation}
This reproduces the insulating Chern-Simons value and the doped Berry-curvature factor used in Ref.~\cite{Sedrakyan2026FiniteFrequency}.

Inside the gap, with $T\ll |m|-|\mu|$, thermally excited carriers only weakly reduce the response:
\begin{align}
\calC_T=\sgn(m)\Bigg[1&-\frac{2T}{|m|}e^{-|m|/T}\cosh\left(\frac{\mu}{T}\right)\nonumber\\
&+O\!\left(\frac{T^2}{m^2}e^{-(|m|-|\mu|)/T}\right)\Bigg].
\label{eq:gap_asymp}
\end{align}
At the band edge, for $m>0$ and $\mu=|m|$,
\begin{equation}
\calC_T=1-\frac{T}{|m|}\ln 2+O\!\left(\frac{T^2}{m^2}\right).
\label{eq:edge_asymp}
\end{equation}
Thus the zero-temperature cusp is rounded over a gate window of order $T$, and the first loss of plateau weight has a universal slope.  In the doped regime, for $|\mu|>|m|$ and $T\ll ||\mu|-|m||$, the Sommerfeld expansion gives
\begin{equation}
\calC_T=\frac{m}{|\mu|}+\frac{\pi^2 m T^2}{3|\mu|^3}+O\!\left(\frac{T^4}{|\mu|^5}\right).
\label{eq:sommerfeld}
\end{equation}
The sign of the leading thermal correction is positive for $m>0$.  This happens because the Berry-curvature weight decreases as $1/\eps^2$: thermal smearing removes more occupied weight below the Fermi level than it adds above it.

\section{Spin-valley structure in graphene}

For a graphene sheet the response must be summed over spin and valley.  Let $\eta_\tau=\pm1$ denote the orientation of valley $\tau$, and let $\zeta_\tau=\pm1$ denote the valley parity of the deformation gauge coupling.  For masses $m_{s\tau}$, the charge response is
\begin{equation}
\calC_T^{\rm gr}(\mu)=\sum_{s,\tau}\eta_\tau\zeta_\tau\,\calC_T(m_{s\tau},\mu,T).
\label{eq:svsum}
\end{equation}
The mixed coefficient in Eqs.~\eqref{eq:rho} and \eqref{eq:j} is obtained by replacing $\calC_T$ with $\calC_T^{\rm gr}$.

For the charge-adding case emphasized in Ref.~\cite{Sedrakyan2026FiniteFrequency}, the mass is a Semenoff mass, $m_{s\tau}=m$, while the deformation gauge coupling is valley odd, $\zeta_\tau=\eta_\tau$.  The four spin-valley species then add:
\begin{equation}
\calC_T^{\rm gr}=4\calC_T(m,\mu).
\label{eq:chargeadding}
\end{equation}
Other combinations of mass and deformation coupling may cancel in the charge channel.  They can still appear in valley, spin, or spin-valley responses.  Equation~\eqref{eq:svsum} is therefore the finite-temperature form of the zero-temperature selection rule.

\section{Driven flexural wave}

We next translate the local response into a mechanical signal.  The weak-deformation gauge field of Ref.~\cite{SedrakyanSinnerZiegler2021} contains, for a pure flexural displacement $h({\bf r},t)$,
\begin{equation}
A^{\rm ph}_a=-\frac{1}{2}\,\partial_a h\,\nabla^2 h,
\label{eq:Aflex}
\end{equation}
up to the microscopic normalization absorbed into $g_{\rm ph}$.  This expression has a simple physical meaning.  A flexural displacement changes the local normal vector of the sheet.  The product of slope, $\partial_a h$, and curvature, $\nabla^2h$, gives a gauge-like field that is even under $h\rightarrow -h$ and quadratic in the deformation amplitude.  Therefore a single sinusoidal mechanical mode does not act as a linear gauge drive.  Its gauge field contains a component at twice the mechanical wave vector and twice the mechanical frequency.

For a traveling wave
\begin{equation}
h(x,t)=h_0\cos(qx-\omega t),
\end{equation}
Eq.~\eqref{eq:Aflex} gives
\begin{equation}
A^{\rm ph}_x=-\frac{q^3h_0^2}{4}\sin[2(qx-\omega t)],\quad A^{\rm ph}_y=0.
\label{eq:Awave}
\end{equation}
The phonon electric field is therefore
\begin{equation}
E^{\rm ph}_x=-\partial_t A_x^{\rm ph}=-\frac{\omega q^3h_0^2}{2}\cos[2(qx-\omega t)].
\end{equation}
The anomaly-induced current is transverse to this phonon electric field:
\begin{equation}
j_y^{2\omega}(x,t)=-\Gamma_T^{\rm gr}\frac{\omega q^3h_0^2}{2}\cos[2(qx-\omega t)],
\label{eq:flex_current}
\end{equation}
where $\Gamma_T^{\rm gr}=(e g_{\rm ph}/4\pi)\calC_T^{\rm gr}$.

Several features of Eq.~\eqref{eq:flex_current} are robust.  The signal is at the second harmonic because the geometric gauge field is quadratic in $h$.  The current is transverse because the parity-odd response has Hall form.  The amplitude changes sign when the Dirac mass changes sign, and it follows the gate-temperature dependence of $\calC_T^{\rm gr}$.  These features do not depend on the microscopic value of $g_{\rm ph}$.  The coupling only fixes the overall scale.

The scaling with the mechanical drive is also distinctive.  The amplitude is proportional to $\omega q^3h_0^2$.  The factor $\omega$ reflects the phonon electric field; a static flexural pattern does not drive a current.  The factor $q^3h_0^2$ reflects the slope-curvature structure of the gauge field.  Long-wavelength bending is therefore weak, while shorter-wavelength ripples or high-curvature suspended modes enhance the response.  This scaling helps distinguish the anomaly-induced current from a deformation-potential background, which couples differently to curvature and need not be purely transverse.

In an open-circuit Hall-bar geometry the transverse anomalous current is balanced by an external electric field.  If $\sigma_{xx}$ is the dissipative longitudinal conductivity and $W$ is the width, a local estimate gives
\begin{equation}
V_\perp^{2\omega}(x,t)\simeq \frac{W}{\sigma_{xx}}j_y^{2\omega}(x,t).
\label{eq:voltage}
\end{equation}
This relation is not universal because contacts, screening, and current redistribution depend on the device.  The universal part is the dependence on mass, gate voltage, and temperature.  A useful measurement would sweep gate voltage at fixed mechanical drive and compare the resulting second-harmonic voltage to the predicted $\calC_T^{\rm gr}(\mu,T)$ line shape.  Repeating the sweep for opposite sublattice mass, or for opposite orientation of the drive, gives a direct sign check.

The local result also states when the simple formula should fail.  If $\omega$ or $vq$ becomes comparable to the gap or to the distance from the chemical potential to the band edge, the response is no longer described by a single Hall coefficient.  Then intraband currents, interband absorption, and edge/contact fields can then modify the phase and spatial profile.  In the adiabatic regime, however, Eq.~\eqref{eq:flex_current} gives the leading anomaly-induced signal.

\section{Static ripple mixed with a dynamic phonon}

A second useful protocol is to combine a static ripple with a small dynamic phonon.  This shifts the electrical response from the second harmonic to the fundamental drive frequency.  Let
\begin{equation}
h(x,t)=h_s\cos(Qx)+h_d\cos(qx-\omega t),\quad h_d\ll h_s .
\end{equation}
Keeping the part of $A_x^{\rm ph}$ linear in the dynamic amplitude gives
\begin{align}
A_{x,{\rm mix}}^{\rm ph}=-\frac{h_s h_d}{2}\Big[&Qq^2\sin(Qx)\cos(qx-\omega t)\nonumber\\
&+qQ^2\cos(Qx)\sin(qx-\omega t)\Big].
\label{eq:Amix}
\end{align}
Therefore
\begin{align}
E_{x,{\rm mix}}^{\rm ph}=\frac{\omega h_s h_d}{2}\Big[&Qq^2\sin(Qx)\sin(qx-\omega t)\nonumber\\
&-qQ^2\cos(Qx)\cos(qx-\omega t)\Big],
\label{eq:Emix}
\end{align}
and
\begin{equation}
j_{y,{\rm mix}}^{\omega}=\Gamma_T^{\rm gr}E_{x,{\rm mix}}^{\rm ph}.
\label{eq:jmix}
\end{equation}
The response is now linear in the dynamic amplitude $h_d$, because the static ripple supplies one power of the deformation.  This is often favorable experimentally.  A weak driven mode can be detected at the fundamental frequency, where lock-in techniques are most sensitive, while the static profile controls the spatial form and strength of the gauge field.

The two terms in Eq.~\eqref{eq:Emix} have a transparent origin.  In the first term the slope comes from the static ripple and the curvature comes from the dynamic phonon.  In the second term the dynamic phonon supplies the slope and the static ripple supplies the curvature.  Their relative size is controlled by $Q/q$.  If the static ripple is much smoother than the driven mode, $Q\ll q$, the first term is dominant and the signal scales as $Qq^2$.  If the static ripple is sharper, the second term can dominate and the signal scales as $qQ^2$.  Patterning the static ripple therefore gives a way to tune both the amplitude and phase of the electrical response.

This mixed protocol is also a clean test of the Chern-Simons origin of the signal.  The spatial dependence is fixed by the known static and dynamic displacement fields, while the gate and temperature dependence enters only through $\calC_T^{\rm gr}$.  Thus the same normalized gate sweep should be obtained in the second-harmonic traveling-wave experiment and in the fundamental-frequency mixing experiment.  Agreement between the two would be difficult to explain by a generic piezoelectric, bolometric, or deformation-potential background.  Those backgrounds can depend on local strain, heating, or contact asymmetry, and need not share the same mass sign, transverse direction, and universal thermal rounding.

The mixed geometry also gives useful phase information.  The two contributions in Eq.~\eqref{eq:Emix} are shifted by a quarter period in space and by a quarter period in time.  Measuring the in-phase and quadrature components of the transverse voltage can therefore separate the two geometric pieces.  If the static ripple is known from microscopy or from the device design, the observed phase pattern provides an internal check of the phonon gauge-field form.  In the adiabatic regime, changing temperature or gate voltage should change only the common amplitude factor and not the phase pattern.

Finally, the static ripple need not be a perfectly sinusoidal pattern.  For a general static profile $h_s({\bf r})$ and a small dynamic displacement $h_d({\bf r},t)$, the linear-in-$h_d$ gauge field is obtained by expanding Eq.~\eqref{eq:Aflex}.  Each Fourier component gives the same kind of mixing between slope and curvature.  The anomaly-induced current is then the transverse response to the resulting phonon electric field.  This form is useful for realistic devices, where the static deformation may be set by a substrate, bubbles, wrinkles, or patterned gates.

\section{Numerical evaluation}

We evaluated Eq.~\eqref{eq:C_closed} by adaptive quadrature.  The Fermi functions were written in an overflow-stable form.  The integral was truncated at a large energy cutoff $E_{\rm max}$, and the analytic tail $m/E_{\rm max}$ was added using the fact that the occupation difference tends to unity at high energy.  All plots use $|m|=1$.

\begin{figure}[htbp]
\includegraphics[width=\columnwidth]{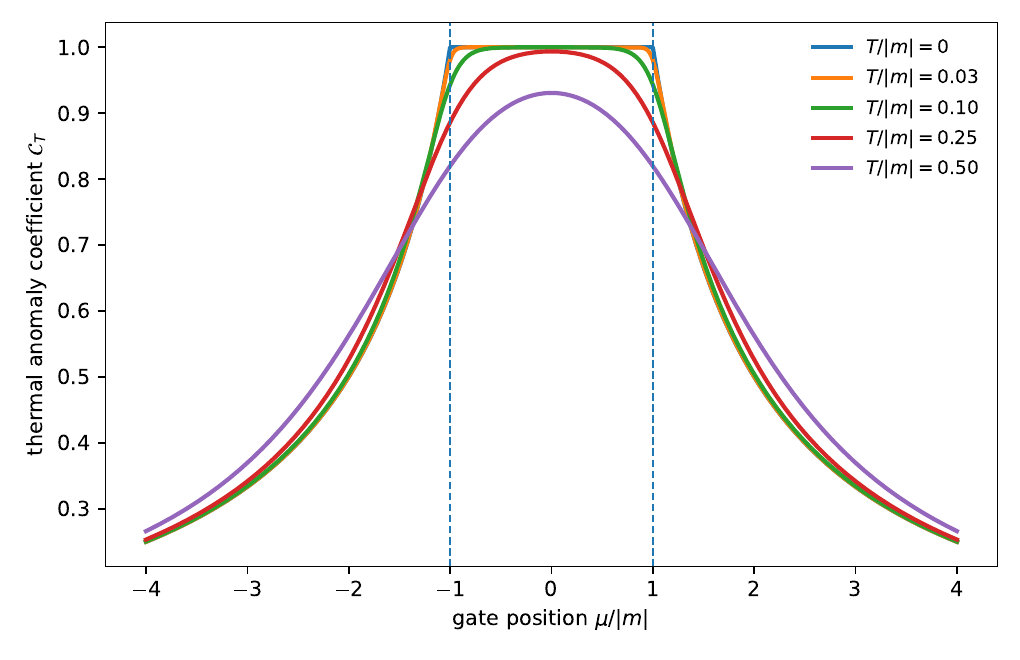}
\caption{Dimensionless one-cone coefficient $\calC_T(m,\mu,T)$ as a function of gate position.  Dashed vertical lines mark $|\mu|=|m|$.  The $T=0$ curve is the plateau-to-$m/|\mu|$ result.  Finite temperature rounds the band-edge nonanalyticity.}
\label{fig:coeff_mu}
\end{figure}

Figure~\ref{fig:coeff_mu} shows the thermal rounding of the zero-temperature plateau.  At small $T/|m|$, the plateau remains nearly flat until the chemical potential approaches the band edge.  At larger temperature the response becomes a smooth peak centered at charge neutrality.  For the charge-adding graphene case, the measured current amplitude is proportional to four times this one-cone curve.

\begin{figure}[htbp]
\includegraphics[width=\columnwidth]{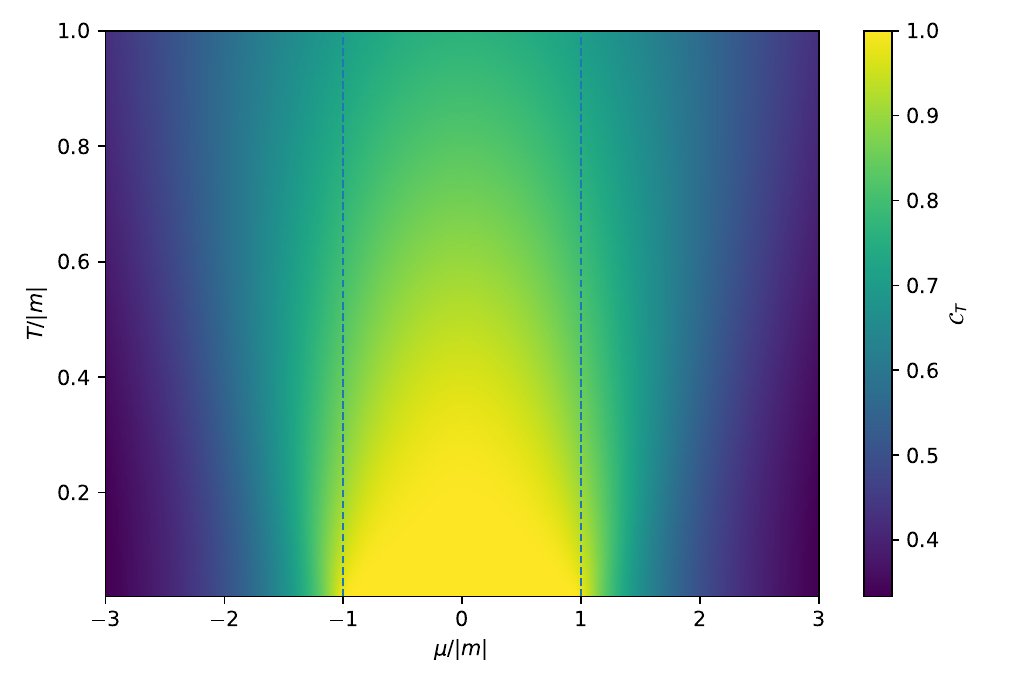}
\caption{Gate-temperature map of $\calC_T$ for $m>0$.  The crossover width near $|\mu|=|m|$ is set by $T$.  The high-doping tail keeps the Berry-curvature scaling.}
\label{fig:heatmap}
\end{figure}

Figure~\ref{fig:heatmap} gives the same information as a gate-temperature map.  This plot is useful for fitting data.  A temperature sweep at fixed gate and a gate sweep at fixed temperature probe the same universal function, up to the overall microscopic prefactor $e g_{\rm ph}$ and the device-dependent voltage conversion.

\begin{figure}[htbp]
\includegraphics[width=\columnwidth]{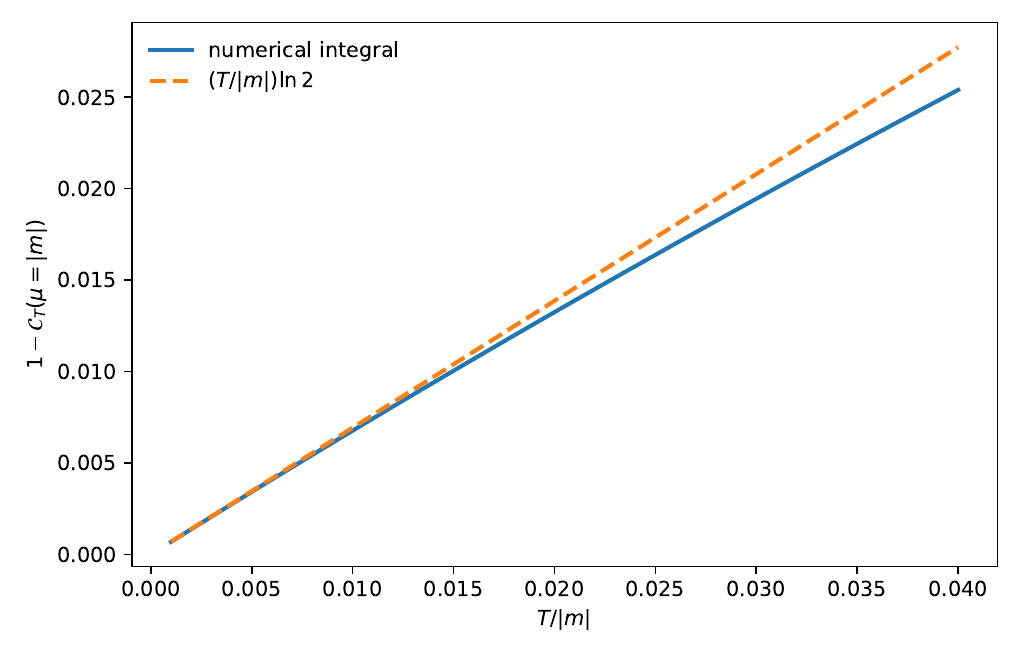}
\caption{Band-edge rounding at $\mu=|m|$.  The initial finite-temperature loss of plateau weight follows Eq.~\eqref{eq:edge_asymp}; deviations at larger $T/|m|$ are higher-order terms.}
\label{fig:edge}
\end{figure}

\begin{figure}[htbp]
\includegraphics[width=\columnwidth]{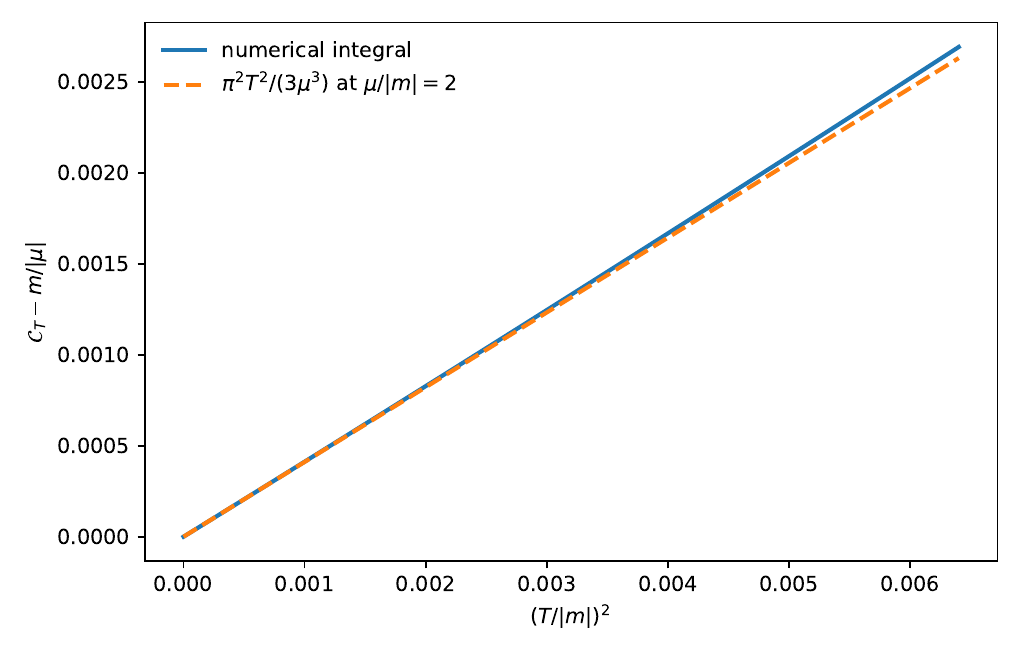}
\caption{Doped-regime Sommerfeld correction at $\mu/|m|=2$.  The difference $\calC_T-m/|\mu|$ is quadratic in $T$ at low temperature and approaches Eq.~\eqref{eq:sommerfeld}.}
\label{fig:asymptotics}
\end{figure}

Figures~\ref{fig:edge} and \ref{fig:asymptotics} test the analytic limits.  At the band edge, the missing plateau weight is initially $(T/|m|)\ln 2$.  In the doped regime, the leading thermal correction is quadratic in $T$ and follows the Sommerfeld coefficient in Eq.~\eqref{eq:sommerfeld}.  These checks probe different pieces of the integral: the band-edge Fermi step and the curvature tail.

\begin{figure}[htbp]
\includegraphics[width=\columnwidth]{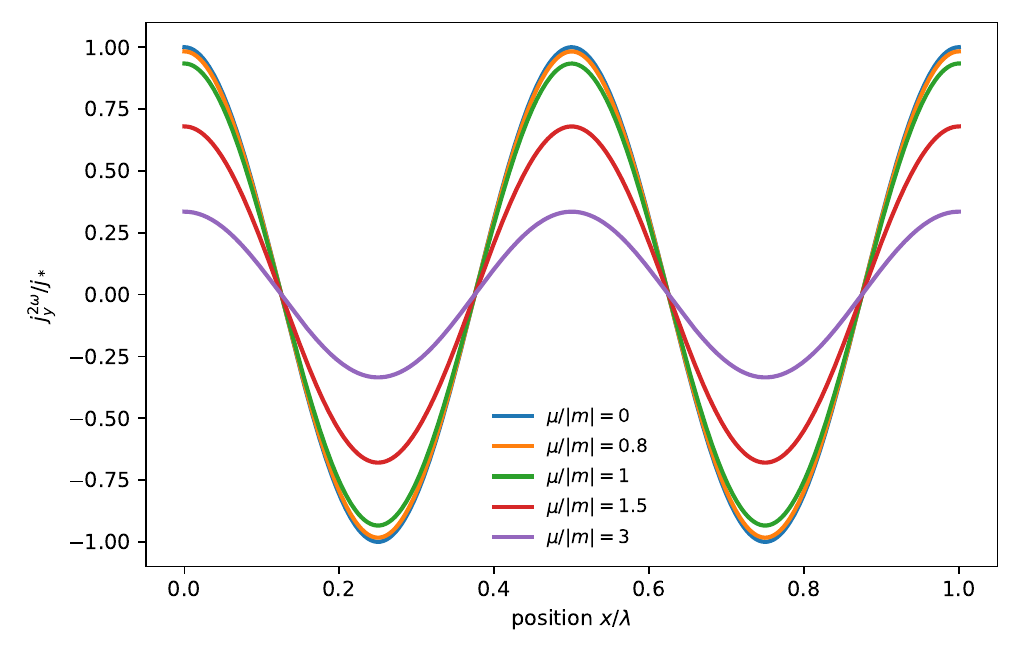}
\caption{Normalized transverse current $j_y^{2\omega}/j_*$ for a flexural wave at $T/|m|=0.12$, where $j_*=(e g_{\rm ph}/4\pi)\omega q^3h_0^2/2$ for one cone.  The charge-adding graphene response multiplies the plotted coefficient by four.  Gate voltage changes the amplitude through $\calC_T$ but does not change the second-harmonic spatial phase.}
\label{fig:flex_profiles}
\end{figure}

Finally, Fig.~\ref{fig:flex_profiles} shows normalized second-harmonic current profiles for a traveling flexural wave.  The spatial shape is fixed by the geometric field in Eq.~\eqref{eq:Awave}.  Changing gate voltage rescales the amplitude through $\calC_T$.  This separation between spatial profile and gate dependence is a strong diagnostic of the anomaly-induced response.

\section{Experimental consequences}

The finite-temperature result leads to several direct experimental tests.

First, the plateau is protected only when the chemical potential is thermally far from the band edge.  A gate sweep should therefore show broadening of order $T$ near $|\mu|=|m|$.  The central plateau should remain almost unchanged until activated carriers become important.

Second, the initial band-edge slope is fixed by $\ln 2/|m|$.  This gives a thermometric check that does not require knowledge of the microscopic phonon-gauge coupling.  The coupling fixes the vertical scale, while the rounded line shape fixes $T/|m|$.

Third, in the doped regime the leading thermal correction increases the magnitude of the response relative to $m/|\mu|$ for $m>0$.  This sign is a useful check.  A simple phenomenological model that only suppresses finite-temperature response would give the wrong trend.

Fourth, because $\calC_T$ is a multiplicative factor in the local regime, the same gate-temperature line shape must appear in the second-harmonic flexural response, in the fundamental-frequency static-ripple mixing response, and in any phonon-flux charge-modulation geometry.  Observing the same $\calC_T$ in more than one geometry would strongly support the mixed Chern-Simons interpretation.

A closely related experimental realization has recently been reported by \cite{Layek2026}, who demonstrated a Hall response induced by dynamic strain in graphene-based heterostructures. In their experiment, oscillatory strain generates a valley-dependent pseudo-electric field which couples to the Berry curvature and produces a measurable transverse charge response, including a Hall signal in the absence of an externally applied electric field. Although the microscopic setting differs from the flexural deformation considered here, the underlying mechanism is closely related: a time-dependent deformation produces a valley-odd effective gauge field whose coupling to Berry curvature yields contributions from the two valleys that add in the charge channel. This experiment therefore provides a direct proof of principle for detecting electromechanical Hall responses generated by strain-induced gauge fields and suggests that the flexural-wave and static-ripple protocols proposed above are experimentally accessible.

\section{Discussion}

The calculation above isolates the local Hall coefficient of a massive Dirac cone at finite temperature.  It does not compute the full finite-$\omega$, finite-${\bf q}$ current-current tensor.  That larger problem contains intraband and interband continua, dissipative terms, and tensor structures tied to the rest frame of the Fermi sea.  Those effects are important outside the adiabatic regime.  They are not needed to describe the leading low-frequency anomaly-induced current.

The main result is Eq.~\eqref{eq:C_closed}.  It replaces the zero-temperature coefficient by a single finite-temperature scaling function.  In the gap the response is exponentially close to the Chern-Simons value.  At the band edge it has a universal thermal rounding.  In the doped regime it reduces to the Berry-curvature tail with a controlled Sommerfeld correction.  After the spin-valley sum, this function gives the thermal line shape of the charge current in sublattice-gapped graphene with valley-odd deformation gauge coupling.

The result is useful because it separates universal and nonuniversal physics.  The function $\calC_T$ is universal within the local Dirac theory.  The overall scale depends on the microscopic value of $g_{\rm ph}$ and on how a local current is converted into a measured voltage.  This means that a single device may require calibration of the vertical scale, but the shape of a gate-temperature sweep is fixed.  The same shape should appear in different mechanical protocols if the observed signal is controlled by the mixed Chern-Simons response.

Several extensions would refine the comparison with experiments.  Disorder can be included through a broadened spectral function or a self-energy.  Nonlocal corrections can be computed from the full finite-temperature polarization tensor.  A realistic Hall-bar calculation can include screening, contacts, and edge current redistribution.  These effects change the conversion from local current to measured voltage, but they should preserve the universal line shape in the clean adiabatic limit.  The finite-temperature coefficient derived here is therefore the natural starting point for quantitative analysis of electromechanical anomaly signatures in gapped graphene.

\begin{acknowledgments}
The research was supported by Armenian HESC grants 21AG-1C024 and 24FP-1F039.
\end{acknowledgments}

\appendix

\section{Derivation of the finite-temperature integral}

For completeness we give the derivation of Eq.~\eqref{eq:C_closed}.  The mixed Hall coefficient is obtained by replacing one electromagnetic vertex in the ordinary Hall response by a phonon-gauge vertex.  In the local limit, the Kubo formula is equivalent to the Berry-curvature expression \cite{Xiao-2010}
\begin{equation}
\sigma^{\rm mix}_{xy}=e g_{\rm ph}\sum_{n=\pm}\int\frac{\dd^2k}{(2\pi)^2} n_F(E_n-\mu)\Omega_n({\bf k}) .
\end{equation}
With the convention used in the main text,
\begin{equation}
\Omega_-({\bf k})=\frac{m v^2}{2\eps_k^3},\quad \Omega_+({\bf k})=-\frac{m v^2}{2\eps_k^3}.
\end{equation}
Thus
\begin{equation}
\sigma^{\rm mix}_{xy}=e g_{\rm ph}\int\frac{\dd^2k}{(2\pi)^2}\frac{m v^2}{2\eps_k^3}\left[n_F(-\eps_k-\mu)-n_F(\eps_k-\mu)\right].
\end{equation}
Using $\dd^2k=2\pi k\dd k$ and $v^2 k\dd k=\eps\dd\eps$ gives
\begin{equation}
\sigma^{\rm mix}_{xy}=\frac{e g_{\rm ph}}{4\pi}m\int_{|m|}^{\infty}\frac{\dd\eps}{\eps^2}\left[n_F(-\eps-\mu)-n_F(\eps-\mu)\right].
\end{equation}
The identity
\begin{equation}
n_F(-\eps-\mu)-n_F(\eps-\mu)=\frac{\sinh(\eps/T)}{\cosh(\mu/T)+\cosh(\eps/T)}
\end{equation}
then gives Eq.~\eqref{eq:C_closed}.

\section{Asymptotic limits}

In the insulating regime, $|\mu|<|m|$, the occupation difference may be expanded as
\begin{equation}
n_F(-\eps-\mu)-n_F(\eps-\mu)=1-e^{-(\eps+\mu)/T}-e^{-(\eps-\mu)/T}+\cdots .
\end{equation}
The correction is dominated by $\eps=|m|$, which gives Eq.~\eqref{eq:gap_asymp}.  At $\mu=|m|$, the leading missing weight is instead
\begin{align}
|m|\int_{|m|}^{\infty}\frac{\dd\eps}{\eps^2}\frac{1}{e^{(\eps-|m|)/T}+1}
&=\frac{T}{|m|}\int_0^\infty\frac{\dd x}{e^x+1}\nonumber\\
&\quad+O(T^2/m^2),
\end{align}
which yields Eq.~\eqref{eq:edge_asymp}.  In the doped regime, for $\mu>|m|$, Eq.~\eqref{eq:C_FD} may be written as
\begin{equation}
\calC_T=m\left[\frac{1}{|m|}-\int_{|m|}^{\infty}\frac{\dd\eps}{\eps^2}n_F(\eps-\mu)\right]
\end{equation}
up to exponentially small valence-band corrections.  Applying the Sommerfeld expansion to $g(\eps)=\eps^{-2}$ gives Eq.~\eqref{eq:sommerfeld}.  The result for $\mu<-|m|$ follows by particle-hole symmetry and depends on $|\mu|$.

\end{document}